\documentclass[aip,jcp,numerical,reprint]{revtex4-1}
\usepackage[utf8]{inputenc}
\usepackage[english]{babel}
\usepackage{amsmath}
\usepackage{amsfonts}
\usepackage{amssymb}
\usepackage{graphicx}

\begin{document}


\title{Hydrogen bonding in the protic ionic liquid triethylammonium nitrate explored by density functional tight binding simulations } 



\author{Tobias Zentel}
\affiliation{Institute  of Physics, University of Rostock, Albert-Einstein-Str.~23-24, 18059 Rostock, Germany}

\author{Oliver K\"uhn}
\email{oliver.kuehn@uni-rostock.de}
\affiliation{Institute  of Physics, University of Rostock, Albert-Einstein-Str.~23-24, 18059 Rostock, Germany}


\date{\today}

\begin{abstract}
The applicability of the density functional based tight binding (DFTB) method to the description of hydrogen bond  dynamics and infrared spectroscopy is addressed for the exemplary  protic ionic liquid triethylammonium nitrate. Potential energy curves for proton transfer in gas and liquid phase are shown to be comparable to high level coupled cluster theory in the thermally accessible range of bond lengths. Geometric correlations in the hydrogen bond dynamics are analyzed for a cluster of six ion pairs. Comparing  DFTB and regular DFT data lends further support for the reliability of the DFTB method. Therefore, DFTB bulk simulations are performed to quantify the extent of geometric correlations in terms of Pauling's bond order model. Further, infrared (IR) absorption spectra are obtained and analyzed putting emphasis on the signatures of hydrogen bonding in the NH-stretching and far IR hydrogen bond range.
\end{abstract}


\maketitle  
\section{Introduction}

Ionic liquids (IL) are discussed as promising candidates for a wide variety of novel applications.~\cite{rogers03_792,douglasr.macfarlane14_232} Their unique physico-chemical properties  derive from the interplay of different types of intermolecular interactions. The ionic character of the molecules leads to strong Coulomb forces, but also dispersion forces and hydrogen bond (HB) interactions are present.~\cite{Greaves08_206,roth12_105026,Ludwig15_13790,hunt15_1257} The subtle balance between these interactions has been reported to lead to situations, where even clusters of like-charge ions are stabilized.~\cite{knorr16_458} For the special case of protic ILs (PILs), the HBs link ions of opposite charge into networks. If the cation is alkylammonium, for instance,  the dimensionality of this network can be tuned by the number of alkyl chains.~\cite{fumino09_3184}  Due to their particular strong HBs, PILs may exhibit proton transfer from the cation to the anion, thus creating a neutral pair of molecules. This may perturb the network, leading to long-range correlated dynamics.~\cite{zentel16_}

Hydrogen bonding manifests itself in  IR spectral features.~\cite{giese06_211,nibbering07_619}  In the far IR region, Fumino \textit{et al.}~\cite{fumino15_2792} could show for a variety of ILs including those with alkylammonium cations, that  signatures of HBs can  be distinguished from contributions due to dispersion forces and intramolecular modes. In the mid IR region,  H-bonding commonly results in a red shift of the involved CH- or NH-stretching vibrational modes with respect to the free vibrations.  However, the spectral signatures could be obscured by the presence of Fermi resonances. This has been illustrated, for instance, for the non-protic IL [C$_n$mim][NTf$_2$], where three different CH$\cdots$N HBs can be distinguished, one of them being influenced by a Fermi resonance with the overtone of a deformation vibration of the imidazolium ring.~\cite{roth12_105026} H-bonding and Fermi-resonances are signatures of the anharmonicity of the potential energy surface for nuclear motion and, therefore, they also have an effect on the width of the absorption bands. In fact the different strengths of the HBs in [C$_n$mim][NTf$_2$] have been shown to give rise to different dephasing times and thus band widths.~\cite{chatzipapadopoulos15_2519}

Computer simulations can provide valuable insights into the structure and dynamics of ILs.~\cite{kirchner15_202,Wang07_1193,Maginn09_373101} For example, density functional theory (DFT) has been used to study the effect of alkyl chain length \cite{bodo12_13878} and environmental contributions \cite{mangialardo14_63} on  IL gas phase cluster structures. An account of crystal effects on the vibrational density of states related to H-bonding has been provided in Ref.~\citenum{mondal15_1994} for [n-alkylammonium][Br].  Computationally more demanding bulk structures of PILs were investigated by Kirchner \textit{et al.}~\cite{zahn10_124506} These authors employed Car-Parinello molecular dynamics (MD)  with periodic boundary conditions to investigate the HB network properties in monoethylammonium nitrat. Inspecting radial, angular and spatial distribution functions, preferred ion orientations and HB properties have been studied and  ion caging has been observed. The latter is often discussed as general feature of IL dynamics.~\cite{Schroder11_024502}

Based on MD trajectory data, power and infrared (IR) spectra can be obtained from correlation functions of  the velocity and the dipole moment, respectively.~\cite{Thomas13_6608}  The quality of these spectra, of course, depends strongly on the underlying potential energy surface driving the dynamics. In Ref.~\citenum{Wendler12_1570} computationally expensive Car-Parinello short-time (tens of ps) MD simulations of  up to 30 ion pairs were performed to obtain power spectra of various imidazolium-based ILs. 
 
\begin{figure}
\includegraphics[width=0.4\columnwidth, angle=-90]{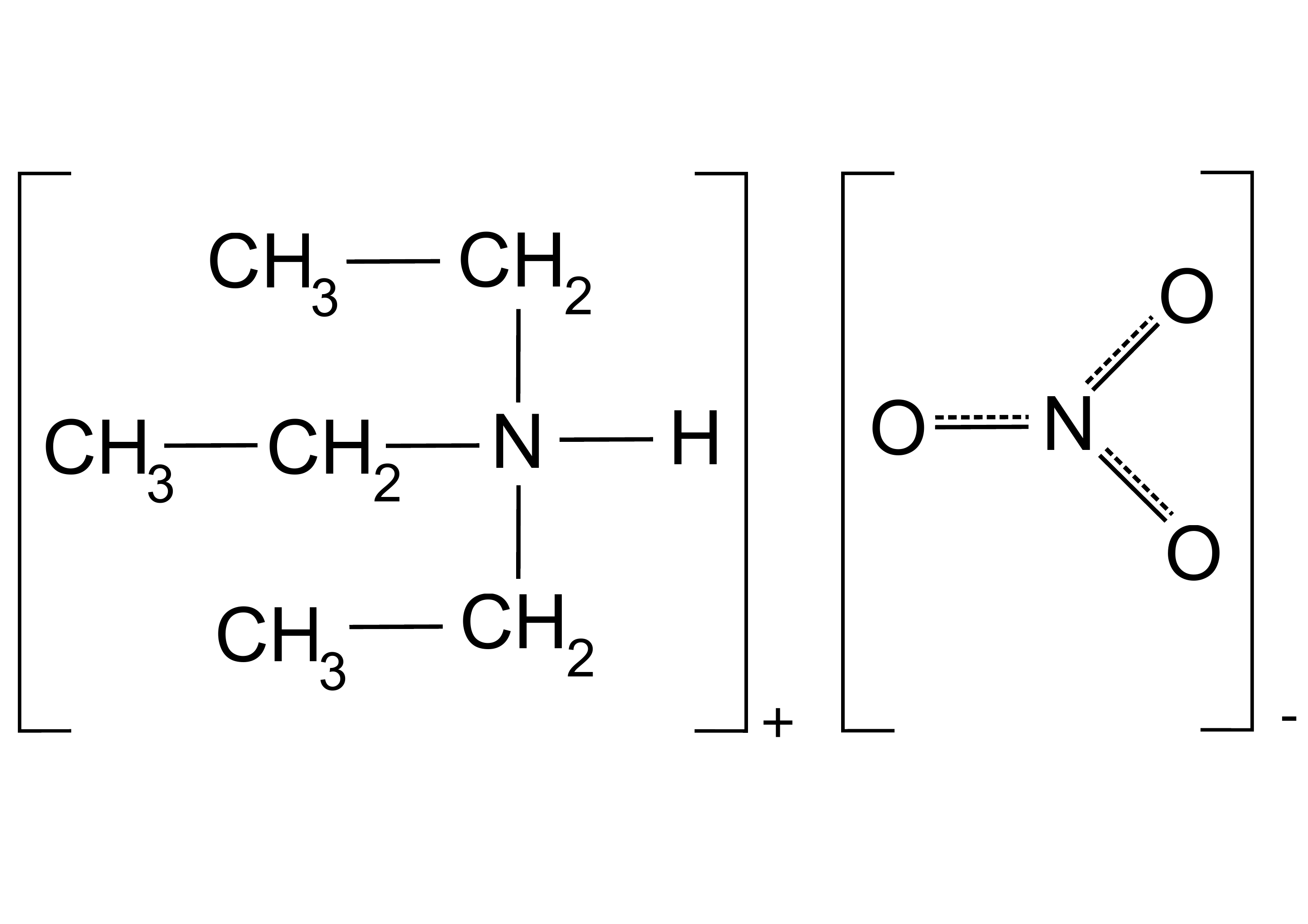}
\caption{Sum formula of the protic ionic liquid forming triethylammonium nitrate (tEAN).}
\label{fig:dipoletypes}
\end{figure}

To reduce the computational costs such as to make nanosecond simulation times accessible, force fields (FFs) with parametrized interactions are usually applied in classical MD simulations.~\cite{Maginn09_373101,koddermann07_2464}  Important properties of molecular systems can be highlighted through FF, for example, the solubility of small molecules in ILs.~\cite{Kerle09_12727} However, the FF parameters are commonly tuned the reproduce thermodynamics properties and thus might not yield reliable  IR spectra. This is due to the fact that the atomic interactions are mostly approximated as harmonic oscillators, that can not describe genuinely anharmonic effects, like the red shift of frequencies in case of HBs. Further, in simple FFs the partial charges do not react to the environment, which can be partly overcome by polarizable FFs.~\cite{Schroder12_3089}  Still, the harmonic or low-order anharmonic approximations applied to bond motions cannot account for reactions, e.g., the proton transfer in PILs. Therefore, it would be beneficial to find methods that provide a reliable  description of the potential energy surface of PILs, such that IR spectra of HBs and proton transfer dynamics can be simulated under bulk conditions.
Here, tight-binding methods might fill the gap between \emph{ab initio} methods and classical MD, one of the flavors being density functional based tight binding theory (for a review, see Ref.~\citenum{seifert12_456}). The self-consistent charge density functional based tight binding method~\cite{elstner98_7260,koskinen09_237} (DFTB) was successfully applied to a large class of problems, including HB interactions in nucleic acid base pairs~\cite{Elstner01_5149}  and PILs.~\cite{Addicoat14_4633} In the latter case it was  shown, for example,  to reproduce experimental bulk structures and gas phase proton affinities from higher level theory of various PILs. Further, correlated fluctuations of ion pairs have been studied for the prototypical PIL triethylammonium nitrate (tEAN), cf. Fig.~\ref{fig:dipoletypes},  in Ref.~\citenum{zentel16_}.  Interestingly, there seem to be no applications of DFTB to IR spectroscopy of bulk ILs to far. This is surprising  since DFTB offers principal advantages with respect to common \emph{ab initio} MD methods. For instance, because atom-centered charges are calculated self consistently on the fly,  no charge localization procedure, such as Voronoi tessellation, is needed to obtain IR spectra, as it would be the case in \emph{ab initio} MD simulations.~\cite{Thomas15_3207} Additionally, no rescaling of the frequencies as a consequence of a  fictitious electron mass as used  in Car-Parinello MD is needed.~\cite{Wendler12_1570} However, these points are of any advantage only if DFTB provides a reasonable description of such IL properties. 

In the present contribution we address the performance of DFTB for the description of HB dynamics and IR spectroscopy of the exemplary PIL tEAN. After introducing the simulation methods in Sec.~\ref{sec:methods}, potential energy curves for proton transfer in gas and liquid phase are compared in Sec.~\ref{sec:pt}  for various methods including high-level coupled cluster theory. DFTB is shown to provide a reliable potential in the vicinity of the ion pair minimum configuration and it is capable of capturing the effect of electronic polarization of the surrounding liquid (Sec.~\ref{sec:ct}). The formation of a  HB comes along with  correlations of the HB geometry, which is investigated in Sec.~\ref{sec:corr}. IR spectra are analyzed in Sec.~\ref{sec:ir}, putting emphasis on the signatures of H-bonding. The results are summarized in Sec.~\ref{sec:sum}.
\section{Simulation Details} 
\label{sec:methods}
\subsection{Molecular Dynamics: Bulk}
A box consisting of 32 tEAN ion pairs with periodic boundary conditions has been simulated using Gromacs 4.5.5.~\cite{hess08_435} Parameters for the FF were taken from OPLS-AA~\cite{rizzo99_4827} and its extension to ILs, especially partial charges were recalculated, by  P\'adua and coworkers.~\cite{canongialopes04_16893,canongialopes04_2038}  To simulate the liquid phase the temperature is set to $177^\circ \mathrm{C} $, well above the melting point of $113-114^\circ$C.~\cite{Greaves08_206} No values for the density are available from literature. In order to obtain the density  and equilibrate the starting geometry a simulation in the NPT ensemble, employing the Parrinello-Rahman barostat and Nos\'e-Hoover thermostat (chain lengths = 10), was run for 1.5 ns using a time step of 0.1~fs and cut-off radii of 0.9~nm. Long-range  Coulomb forces were evaluated using the particle mesh Ewald summation. The MD simulation gives a density of 1.0486~g/cm$^3$, which compares well with  the density of monoethylammonuim nitrate, 1.216~g/cm$^3$, at $27^\circ \mathrm{C}$.~\cite{Greaves08_206} The addition of alkyl chains is expected to lower the density and FF simulations are known to reproduce liquid densities well.~\cite{Maginn09_373101}  The obtained box is used as an input for a MD run in the canonical ensemble for 5~ps, using a time step  of 0.5~fs and the same temperature as before.  Starting from the  NVT output, initial conditions for microcanonical NVE trajectories are sampled along the canonical trajectory and run using the same FF.

In a second setup, the final FF  NVT output is used as a starting structure for DFTB simulations. First, the structure is equilibrated in a NVT ensemble for 5~ps from which five initial structures for NVE trajectories are drawn randomly. All microcanonical trajectories were simulated up to a length of 25~ps using a time step of 0.5~fs. DFTB simulations were performed with the DFTB+ code \cite{aradi07_5678}, including the  3rd order correction\cite{yang07_10861} and  employing the Slater-Koster parameters of the 3ob set.~\cite{gaus13_338} Van der Waals dispersion forces are included \textit{a posteriori }in the DFTB calculations \cite{Elstner01_5149}, with parameters taken from the Universal Force Field. The system was simulated at the $\Gamma$-point only. One reference calculation with Monkhorst-Pack~\cite{monkhorst76_5188} 2x2x2  $k$-point sampling  was performed.
\subsection{Ab Initio Molecular Dynamics: Hexamer Cluster} 
A DFT optimized \cite{kastner09_11856} hexamer gas phase cluster structure was used as an input for a MD trajectory propagated using Terachem,  version 1.5K.~\cite{ufimtsev09_2616} The BLYP exchange-correlation functional was used together with the 6-31+G* basis set.  Dispersion was accounted for by  Grimme's D3 method.~\cite{grimme10_154104} A 10~ps trajectory  (time step 0.4~fs) has been simulated  using Langevin dynamics with the coupling parameter set to 1~ps and a reference temperature of 300~K. The first 1.2~ps of the trajectory were considered as equilibration period. For comparison, the same initial hexamer structure was simulated with  DFTB using  the same setup as for the bulk simulations.

\subsection{IR Spectra}
IR spectroscopy probes the dynamics of the fluctuating dipoles and hence contains information about H-bonding between ion pairs. Here, the molecular dipoles were calculated from the atomic charges positioned at the  nuclei, \emph{i.e.}\ either the Mulliken (DFTB) or the  partial charges from the FF. The IR spectrum, $I(\omega)$, was obtained by Fourier transformation  of the dipole trajectory, $\mu(t)$, which was calculated as the sum of the individual molecular dipoles in the box. A  Kaiser window function, $\kappa(t)$,  with Kaiser parameter 14 was employed. Thus the spectrum is given by 
 \begin{equation}
I(\omega) = \omega^2 \langle | \int\limits_{0}^{T} \exp(-i \omega t)   \vec{\mu}(t) \kappa(t)   \mathrm{d}t \ |^2 \rangle \, .
\label{eq:IR}
\end{equation}
Here, the averaging is taken with respect to the NVE trajectories.
The spectra discussed below were obtained by moving averages of 21 steps ($ \widehat{=} \ 26.7~\mathrm{cm^{-1}}$) to smooth the curves.  To disentangle the spectral contributions, ions that are fixed as pairs for more than 15~ps were identified as `fixed' bound pairs, while all others are called  `flexible' bound pairs. Spectra have been calculated separately for both motifs. Moreover,  the dipole moments of the NH-bonds and CH$_3$ groups were extracted to give their contribution to the total spectrum.

\subsection{Potential Energy Curves} 
For a selected ion pair from the DFTB bulk box, having a typical geometry, the NH-distance, $r_{\mathrm{NH}}$, has been varied in $0.1 \ \mathrm{\AA}$ steps in the range of $1$ to $2.2 \ \mathrm{\AA}$ along the vector associated with the NH-bond. All other positions have been fixed at the snapshot geometry. For each case single point energy calculations  were performed using the FF, DFTB,  DFT, and Coupled Cluster theory (CCSD(T)).  The  Turbomole 6.5 software package \cite{turbomole6.5,ahlrichs89_165,hatting02_2111,hatting00_5154,treutler95_346} was used to calculate DFT  (B3LYP,  def-SV(P) basis set)  and  CCSD(T) (def2-TZVP basis set) energies.~\cite{hatting02_2111,hatting00_5154} Gromacs and DFTB+ were used for FF and DFTB energy calculations, respectively. In these calculations only the selected pair was treated explicitly. Thereby, gas and liquid phases were compared. In the latter case, all surrounding molecules enter via their classical atom-centered point charges for a given bond length, which were taken to be the  DFTB Mulliken charges.
\section{Results} 
\subsection{Proton Transfer Reaction} 
\label{sec:pt}
\begin{figure}
\includegraphics[width=\columnwidth]{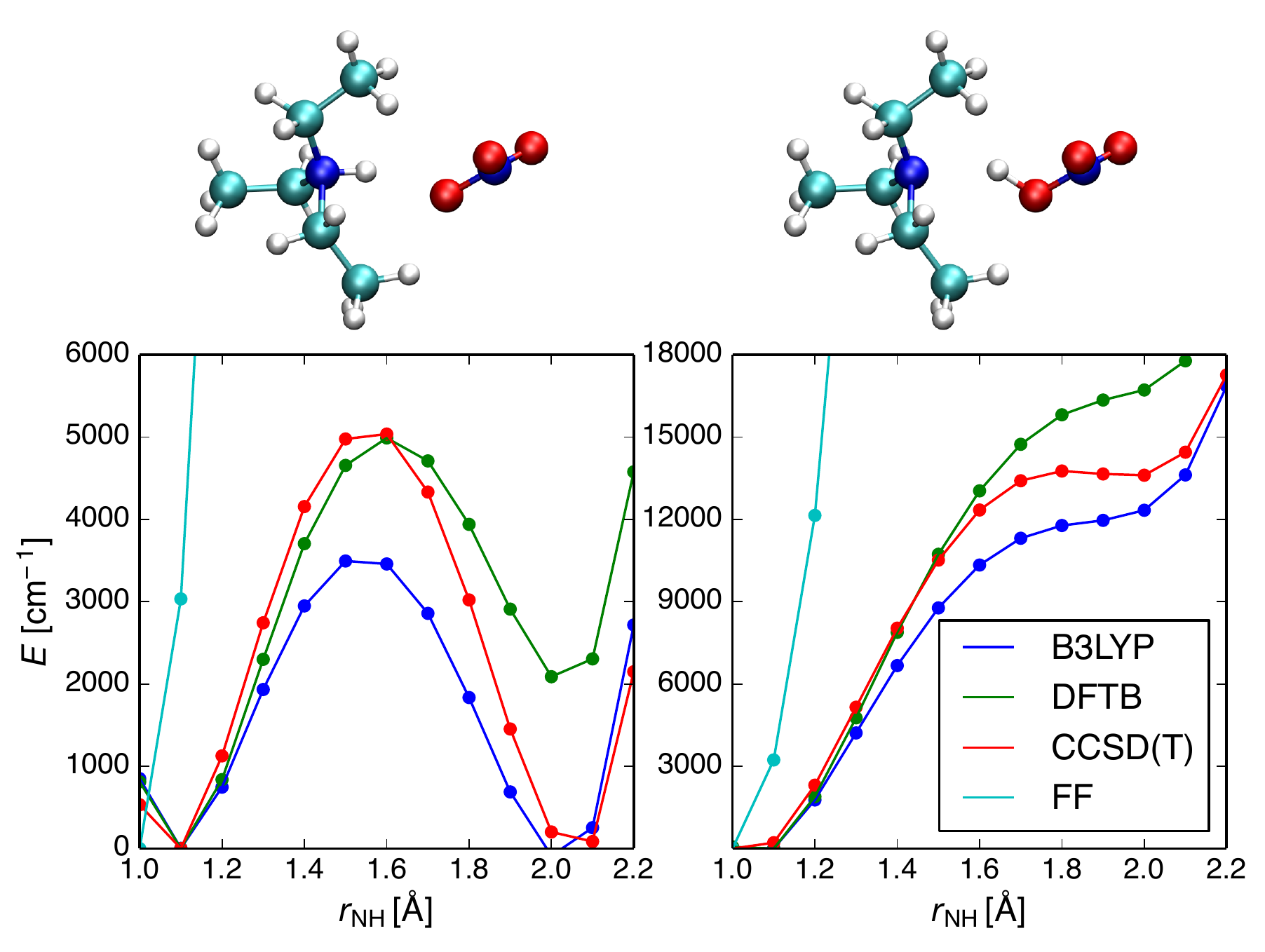}
\caption{ Top: tEAN as an ionic (left) and neutral (right) pair. Bottom: Potential energy curve along the NH-bond vector calculated with different methods for an ion pair in gas phase (left) and an ion pair surrounded by classical point charges derived from DFTB Mulliken charges (right). }
\label{fig:potential}
\end{figure}

To investigate the type of HB and to  compare the performance of different quantum chemical methods, the potential energy curves along the NH-bond are shown in Fig.~\ref{fig:potential} for gas and liquid phase. In  the following discussion special emphasis is laid on the performance of DFTB with respect to DFT and CCSD(T).  Note that the nuclear reorganization energy is not considered here, \emph{i.e.}\ the resulting potential is the one seen by the proton during instantaneous transfer. Electronic polarization is accounted for in the liquid phase case due the NH-bond length dependent Mulliken charges of the surrounding. 

In gas phase, the potential shows a double well character for all methods except FF, see left panel of Fig.~\ref{fig:potential}. The  barrier height is 3500~$\mathrm{cm^{-1}}$ for B3LYP, which is known to underestimate HB barrier heights \cite{elstner06_316}, and 5000~$\mathrm{cm^{-1}}$ for DFTB and CCSD(T). While the DFTB and CCSD(T) barrier heights match, the minimum corresponding the the neutral pair is about 2000~$\mathrm{cm^{-1}}$ deeper for CCSD(T) and B3LYP than for DFTB. In fact DFT/B3LYP even predicts the neutral pair to be more stable than the ion pair. In strong contrast, FF energies show the expected harmonic behavior and, of course, cannot describe proton  transfer. 

The potential energy curves change dramatically upon placing classical point charges around the ion pair. The second minimum reduces to a shoulder in DFTB and DFT/B3LYP and to a very shallow minimum in CCSD(T). The barrier increases to $\approx 13800~\mathrm{cm^{-1}}$ in the CCSD(T) calculation. This finding fits well to the argument by  Hunt \textit{et al.}~\cite{hunt15_1257}, \textit{i.e.}\ that in polar solvents charged ion pairs are stabilized. The energy of the neutral pair must thus be higher than that of the ion pair, resulting in an almost disappearance  of the second minimum if the latter is embedded into point charges. The energetics from DFTB matches that from CCSD(T) for bond lengths $r_{\mathrm{NH}} \leq 1.6~\mathrm{\AA}$, but DFTB yields  energies higher than DFT and CCSD(T) in the region of the second minima. In passing we note that the potential for a full quantum mechanical DFTB calculation as used in the MD simulations below shows a barrier of  $\approx 11480~\mathrm{cm^{-1}}$ and a second minimum at $\approx 11410~\mathrm{cm^{-1}}$.

IR intensities are determined by the dipole gradient along the considered vibrational coordinates. In order to check the performance of DFTB, we have calculated the dipole gradients projected along the N-H bond
 for the liquid phase configuration of Fig.~\ref{fig:potential}. This gave 2.18~D/\AA{} and 2.21~D/\AA{} for DFTB and DFT/BLYP, respectively. 
%
Summing up,  DFTB with third-order correction yields rather good potential energy curves for NH-bond motion in the liquid phase if compared to CCSD(T).  Notice that the third-order correction is essential. Retaining only the second-order term  and using the mio parameter set~\cite{elstner98_7260} results in a rather different potential profile (not shown). Further, dipole gradients for NH- and CH-stretching vibrations are comparable with DFT/B3LYP.
\subsection{Charge Redistribution Upon Proton Transfer}
\label{sec:ct}
\begin{figure}
\includegraphics[width=\columnwidth]{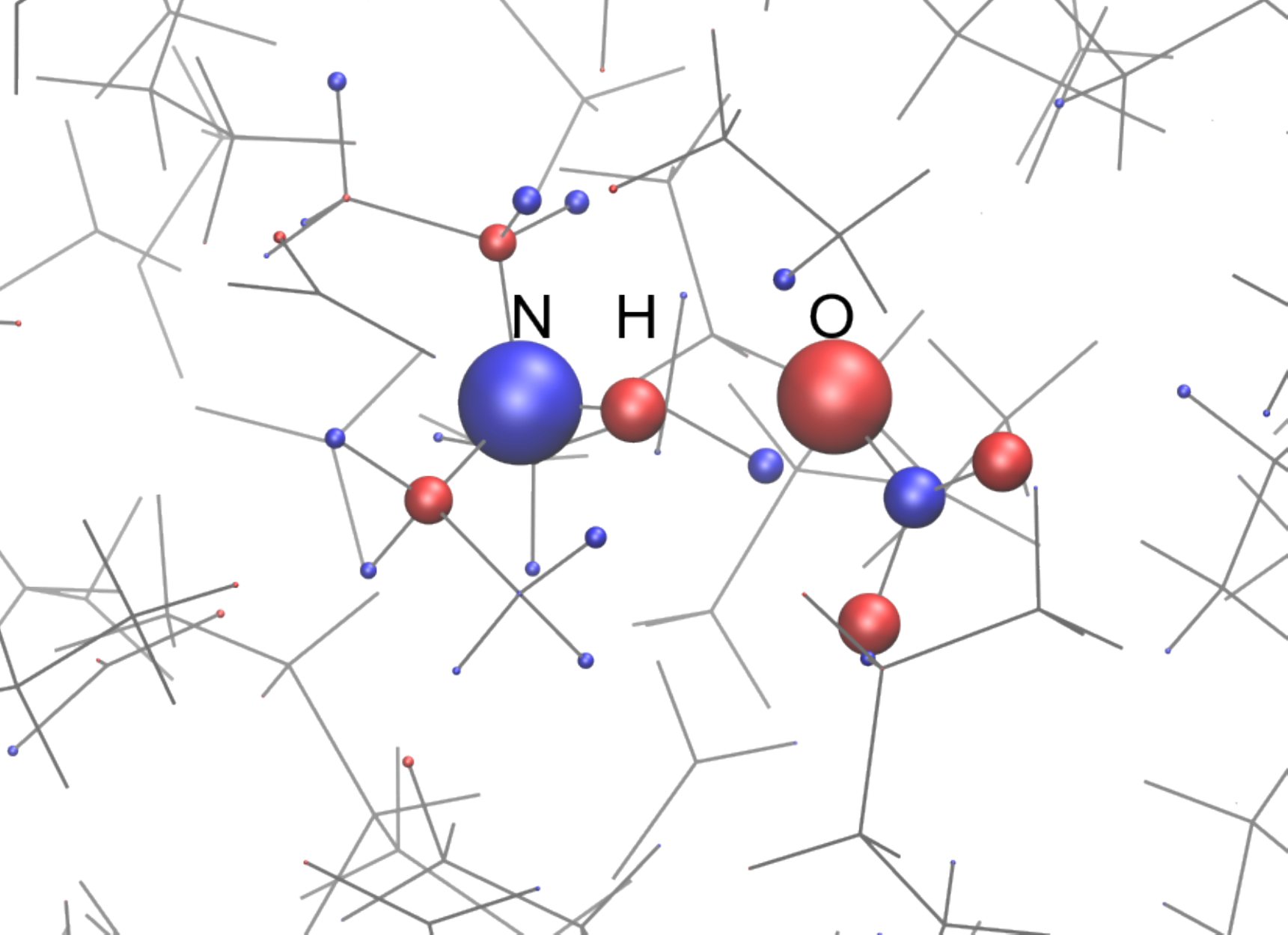}
\caption{Difference of the Mulliken charges between the snapshots at $r_{\mathrm{NH}} = 1.1~\mathrm{\AA}$ and  $r_{\mathrm{NH}} = 2.0~\mathrm{\AA}$ (blue negative, red positive). The absolute values of charges differences are encoded by the size of the spheres. For visibility reasons the sizes of the spheres for N and O atoms of the central pair are scaled by a factor of $0.4$.  }
\label{fig:chargediff}
\end{figure}
An important aspect for understanding the overall mobility of the ions and the observed conductivities is the charge transfer between the ion pairs, as was recently highlighted by Holloczki \textit{et al}.~\cite{holloczki14_16880} Depending on the applied charge localization method, the charges at the ions for various ILs can be considerably lower than $\pm 1e$. Using Mulliken charge localization on a set of ILs, the individual ions carried absolute charges between $0.88e$ and $0.95e$.~\cite{holloczki14_16880} Thus, any method used to simulate ILs should incorporate these effects. In the present DFTB bulk box the average charge of a $\mathrm{NO_3^-}$ anion    is $|q|=0.95e$, which is in the range of the observed charges in Ref.~\citenum{holloczki14_16880}. 

In order to investigate how the charges of the environment react to the proton transfer in the selected pair of the previous section, the difference in DFTB Mulliken charges between the ion and neutral pair is plotted in Fig.~\ref{fig:chargediff}.  The largest charge reduction is found on the HB donor N atom  with  $\Delta q =-0.36e$, while the largest charge increase is found on the HB acceptor O atom ($\Delta q =+0.32e$). Furthermore, the transferred proton becomes more positively charged as do the other two oxygen atoms of the NO$_3^-$. In total the charge redistribution is a local effect as the environment, including the alkyl side chains of the cation, does not react appreciably to the charge transfer. 
\subsection{HB Geometry Correlations}
\label{sec:corr}
\begin{figure}
\includegraphics[width=\columnwidth]{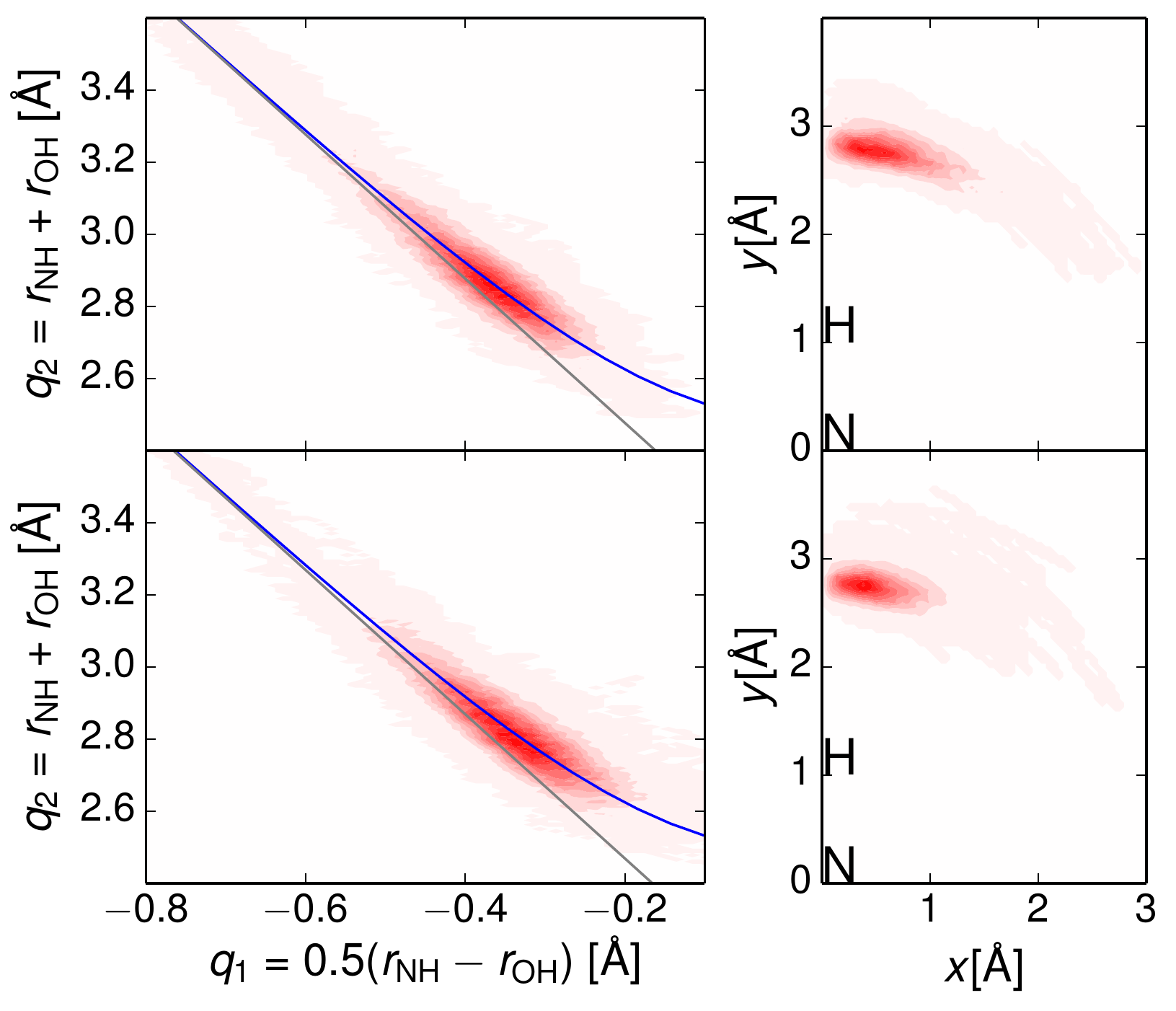}
\caption{Left: HB  spatial correlation plots from DFTB (top) and DFT (bottom) trajectories averaged over all 6 pairs of a gas phase cluster along a 10~ps trajectory.
Right: The position of the oxygen atom in the plane defined by the positions of the N, H and O atom. The N-H bond defines the $y$-axis and the origin is set to be the nitrogen position. The approximate hydrogen position is marked as H.}
\label{fig:HBgeocluster}
\end{figure}

\begin{figure}
\includegraphics[width=\columnwidth]{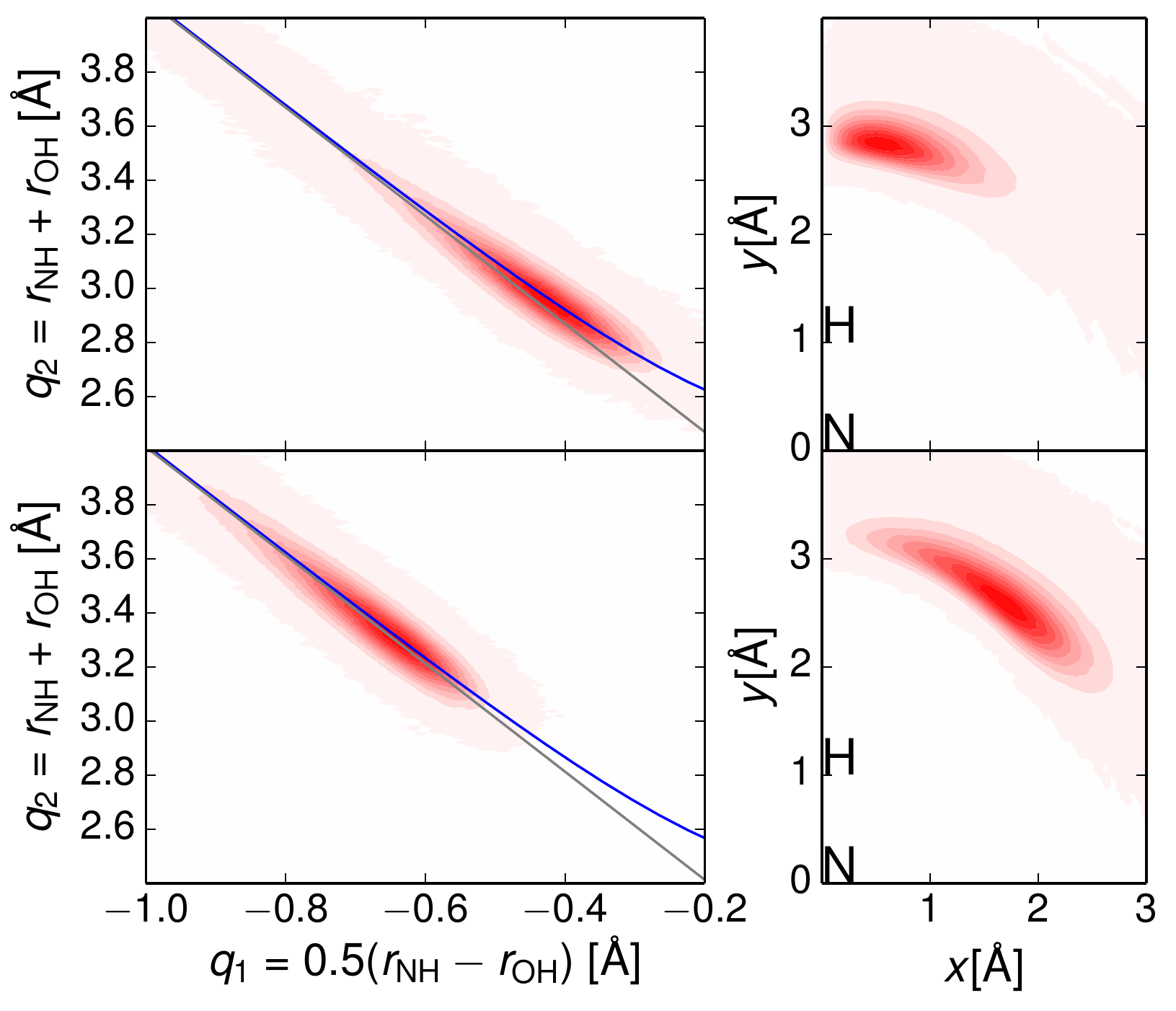}
\caption{Left: HB spatial correlation plots from DFTB and MM trajectories averaged over all 32 pairs of a box along 25ps trajectory.
Right: The position of the oxygen atom in the plane defined by the positions of the N, H and O atom. The N-H bond defines the y-axis and the origin is set to be the nitrogen position.The approximate hydrogen position is marked as H.}
\label{fig:HBgeobulk}
\end{figure} 

There is considerable evidence for the correlation between the HB length (donor-acceptor distance) and the position of the H atom or proton within the HB.~\cite{limbach04_5195,limbach06_193,yan10_15695} In the following we will compare correlation plots obtained for DFTB and DFT cluster calculations as well as for DFTB and FF bulk results. From the respective MD trajectories the positions of the cationic nitrogen, the nearest anionic oxygen atom and the H-bonding H atom are extracted. This information is used to determine the  distribution of HB coordinates $q_1 = 0.5(r_{\mathrm{NH}} -r_{\mathrm{OH}} ) $ and $q_2 = r_{\mathrm{NH}}  + r_{\mathrm{OH}} $ along the trajectories.~\cite{limbach06_193} Notice that in none of the trajectories proton transfer events have been observed.

Before discussing the $q_1$ \textit{vs}.\ $q_2$ correlation, averaged NO and OH distances are compared for the different setups in Tab.~\ref{tab:HBgeo}. 
First, we note that going from cluster to the bulk box using DFTB the average N-O and O-H distances are increasing, indicating stronger HBs  in the gas phase cluster. Second, the distances also increase when going from  DFT via DFTB to FF simulations. $\left< r_{\mathrm{NO}} \right>/\left< r_{\mathrm{OH}} \right> $ is larger by 2\%/7\% in DFTB cluster simulations compared to BLYP simulations, but increases by 8\%/19\% when using FF instead of DFTB  in the bulk system. The N-H distance shows the expected reverse behavior, it decreases by 5\% when the FF is used instead of DFTB.  
In passing we note that using a Monkhorst-Pack 2x2x2 $k$-point sampling instead of the $\Gamma$ point only, the HB geometry changes only marginally. 
\begin{table}
\begin{tabular}{|l|c|c|c|c|}
\hline 
& cluster DFT  & cluster DFTB &  bulk DFTB  &  bulk FF  \\ 
\hline
\hline
$\left<  \alpha \right>\ [^{\circ}] $ & 18.0 & 20.9 & 25.6 & 40.4 \\ 
\hline 
$\left< r_{\mathrm{NO}} \right> \  [\mathrm{\AA}]$ & 2.80 & 2.85 & 2.97 & 3.20 \\ 
\hline 
$\left< r_{\mathrm{OH}} \right>  \ [\mathrm{\AA}]$ &1.77 & 1.85 & 2.00 & 2.38 \\
\hline
$\left< r_{\mathrm{NH}} \right>  \ [\mathrm{\AA}]$  &1.07 & 1.06 &1.06 &1.01 \\ 
\hline
\end{tabular}
\caption{HB distances and angle for different setups as indicated. The average was performed for all H-bonded pairs in the simulation box.}
\label{tab:HBgeo}
\end{table}

The $q_1$ vs.\ $q_2$ correlations are plotted in Fig.~\ref{fig:HBgeocluster} and \ref{fig:HBgeobulk} (left panels) for the cluster and bulk simulations, respectively. If the distances  $r_{\mathrm{NH}} $ and $r_{\mathrm{OH}}$ were fully uncorrelated, the results would lie on a straight line with the slope -2, plotted for reference as a green line. Instead within the  HB  the H atom is pulled towards the oxygen acceptor once they approach each other, \textit{i.e.}\ $r_{\mathrm{OH}} $ is reduced as $r_{\mathrm{NH}} $ increased. This results in deviations from the linear relation and shifts the curve at a given $q_1$ to larger $q_2$ values. Furthermore, the closer $q_1$ is to zero (\textit{i.e.}\ in the middle of the HB), the stronger is the bond. The limiting symmetric case where  $ r_{\mathrm{NH}}  \approx r_{\mathrm{OH}} $ is usually assumed for very strong barrierless HBs.

The geometric correlations of the HB in the hexamer cluster simulations are plotted in the left panels of Fig.~\ref{fig:HBgeocluster}. Overall, the HB strength in the BLYP (top) is stronger than in DFTB (bottom) simulation, since the $q_1$ values are closer to zero. Comparing the shape of the distributions, DFT predicts a slightly stronger correlation as compared with DFTB, which is of course a consequence of the stronger HB.  In the bulk simulations summarized in the left panels of Fig.~\ref{fig:HBgeobulk}, the average $q_1$ for the FF  is much smaller than the corresponding value for DFTB, indicating a much weaker HB. Again this comes along with a stronger correlation  for the case of DFTB as compared with FF. This is in accord with the average bond distances in Tab.~\ref{tab:HBgeo}. 

To classify the geometric correlations in a more quantitative way, the results are compared to other weak and medium strong HBs. This is done by using  the valence bond model of Pauling. Here,  starting from the  HB distances, $r_{\mathrm{NH}}$ and $r_{\mathrm{OH}}$, bond orders are defined as~\cite{pauling47_542} $p_i =  \exp(-(r_{i}-r^{\mathrm{eq}}_{i})/b_i)$ with $i =\{\mathrm{NH}, \mathrm{OH} \}$. Under the constraint that the sum of the two bond orders must equal one, the two coordinates depend on each other and the  proton transfers can be described by a single coordinate.~\cite{limbach04_5195} This path is drawn in  Figs.~\ref{fig:HBgeocluster} and \ref{fig:HBgeobulk} as a blue line.

The equilibrium distances $r^{\mathrm{eq}}$ refer to isolated molecules without HB interactions and are set to  $r_{\mathrm{NH}}^{\mathrm{DFTB}} = 1.037~\mathrm{\AA}$ and $r_{\mathrm{OH}}^{\mathrm{DFTB}} = 0.982~\mathrm{\AA}$ in the DFTB and FF analysis results and $r_{\mathrm{NH}}^{\mathrm{BLYP}} = 1.034~\mathrm{\AA}$ and $r_{\mathrm{OH}}^{\mathrm{BLYP}} = 0.991~\mathrm{\AA}$ in BLYP. These values were obtained from energy-optimized geometries of the respective individual molecules. The parameters describing the bond order decay are set in all plots to $b_{\mathrm{NH}}=0.351~\mathrm{\AA}$ and  $b_{\mathrm{OH}}=0.321~\mathrm{\AA}$. These values have been obtained by changing the  ones reported for typical medium and strong NHO HBs\cite{steiner98_7041} ($b_{\mathrm{NH}}=0.385~\mathrm{\AA}$ and  $b_{\mathrm{OH}}=0.371~\mathrm{\AA}$) such as to fit the  observed geometric correlations for the DFTB cluster case. The obtained values are larger than the references ones, which have been derived from experimental crystal structure data of non-ionic compounds.~\cite{steiner98_7041}  This indicates somewhat weaker HBs in the present case.

Comparing the blue and green curves with the distributions in Figs.~\ref{fig:HBgeocluster} and \ref{fig:HBgeobulk} lends support for the simple Pauling bond order model. It  also gives further evidence that the HB predicted by the FF is essentially   uncorrelated, while  DFTB (bulk and cluster) and BLYP (cluster) results lie in the correlated region (where blue and green lines differ).

Finally, we address the nonlinearity of the HBs in the different simulations setups. The distribution of the O atom's position in a plane defined by the three atoms NHO is shown in right panels of Figs.~\ref{fig:HBgeocluster} and \ref{fig:HBgeobulk}.  In the DFTB  and BLYP cluster simulations only one relatively fixed HB configuration seems to be present. In the hexamer cluster the average  NHO HB angle is $20.9^{\circ}$ in DFTB and $18.0^{\circ}$ in BLYP, \textit{i.e.}\ both values are rather close to each other (\textit{cf}.\ Tab.~\ref{tab:HBgeo}). In bulk DFTB the average HB angle  is increased to $25.6^{\circ}$ and the distribution is somewhat broader than in the cluster case.  In the FF simulation the O atom's position is very flexible yielding  an average angle of $40.4^{\circ}$. In fact, on average per $10$~ps trajectory a cation changes $6.4$ times its H-bonded anion in DFTB, but  $41$ times in FF.

In summary, comparing bulk DFTB and FF simulations, HB distances are shorter, more linear and pair geometries are less flexible in DFTB than in FF. The current FF fails to describe geometric HB correlations.  In fact due to the poor description of HBs,  the FF trajectory  yields ion pairs, which frequently change their partners.

\subsection{IR spectra}
\label{sec:ir}
\begin{figure}
\includegraphics[width=\columnwidth]{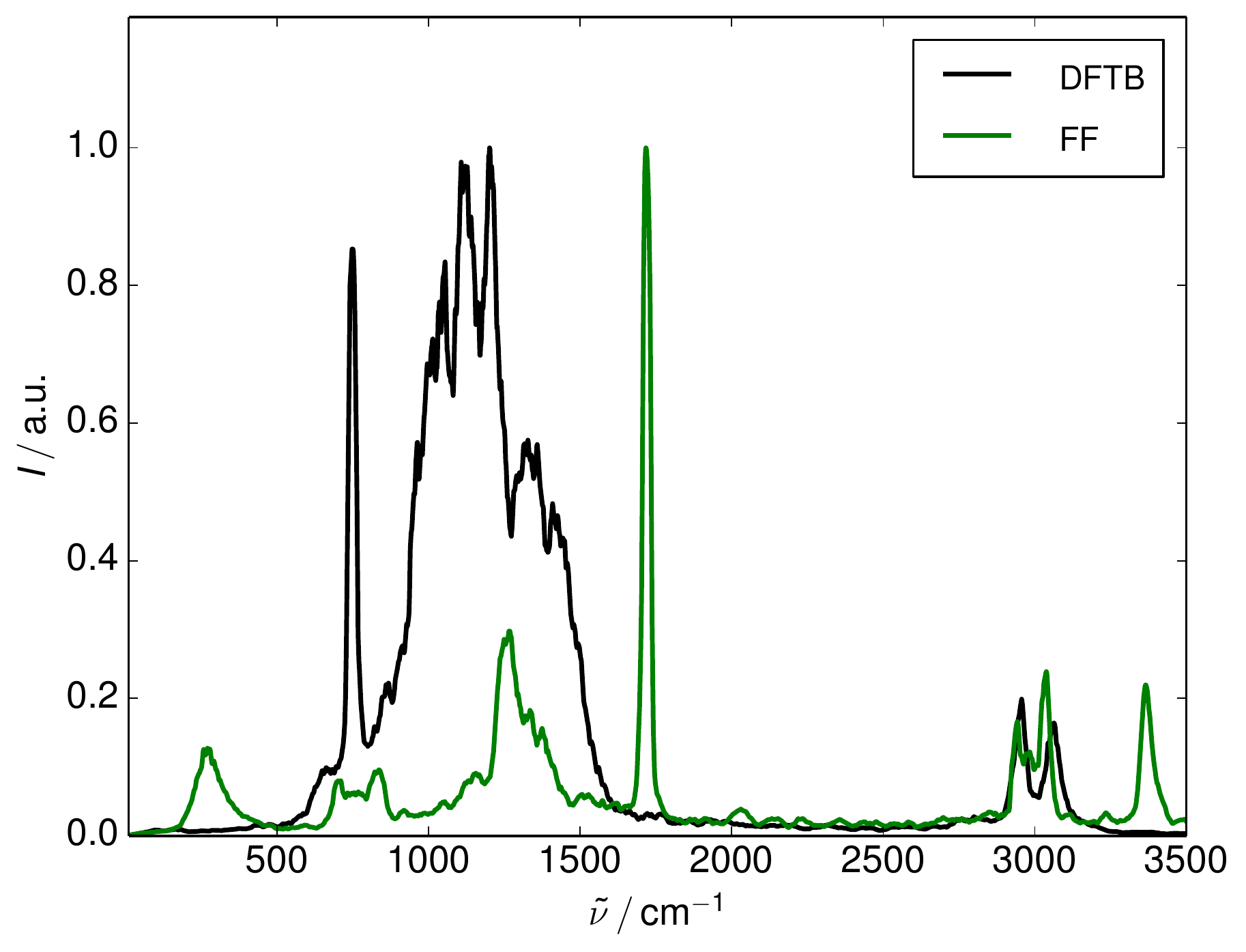}
\caption{IR spectra obtained from the FF and DFTB simulations according to Eq.~\eqref{eq:IR}.}
\label{fig:IR_MMvsDFTB}
\end{figure}

IR spectra were calculated from the DFTB and FF bulk trajectories according to Eq.~\eqref{eq:IR}. A global comparison of the two methods in the investigated  spectral range is show in Fig.~\ref{fig:IR_MMvsDFTB}. Overall, one notices that there are some overlapping regions, but also pronounced peaks predicted by the FF at positions where the DFTB spectrum is essentially flat. For example,  there is a strong peak around 1750~$\mathrm{cm^{-1}}$ in FF, which has no counterpart in DFTB. Further, in the region around 3000 $\mathrm{cm^{-1}}$  both spectra show a double peak structure. However, the NH-stretching vibration in the FF spectrum is located at approximately 3400~$\mathrm{cm^{-1}}$, where DFTB gives no signal. An isolated amine NH-stretching vibration is expected in this region, however due to  H-bonding, this mode should shift to lower frequency, as is the case in the DFTB results. In fact, the absence of higher frequency peaks indicates that all NH-stretching vibrations  are involved in H-bonding. In passing we note that Bodo \textit{et al.}~\cite{bodo12_13878} reported DFT cluster calculations for mEAN, which nicely showed the saturated of the free NH-vibrations with increasing cluster size.   For reasons discussed above the failure of the FF simulation in predicting the IR spectrum does not come unexpectedly. Therefore, in the following only the DFTB spectra will be used for analysis. 
\begin{figure}
\includegraphics[width=\columnwidth]{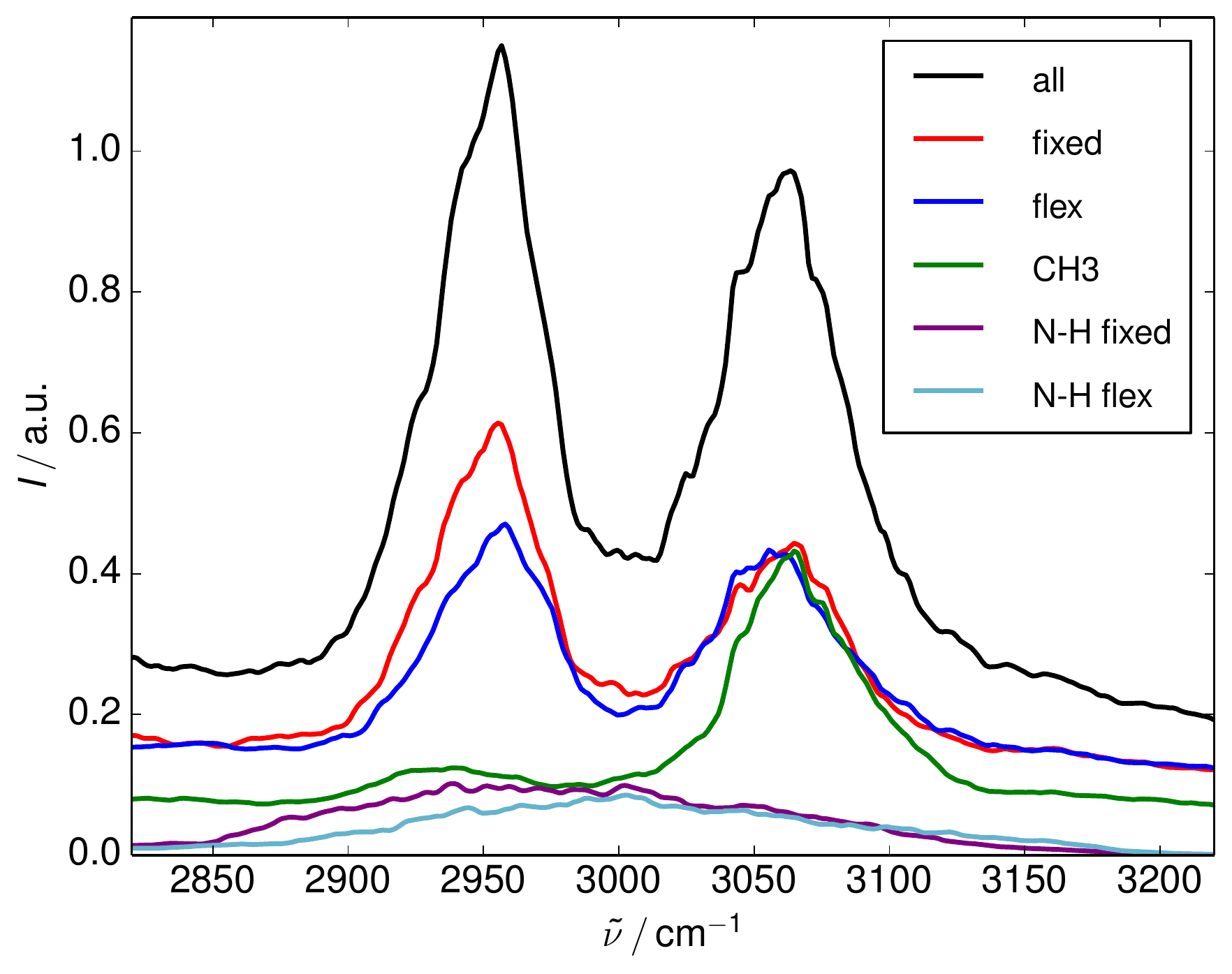}
\caption{DFTB IR spectra in the range of the NH- and CH-stretching  vibrations  according to Eq.~\eqref{eq:IR} (curves are plotted with different offsets for better readability). The separate contributions of fixed and flexible NH$\cdots$O HBs are shown for the whole molecules and for the NH-bond vibrations only. Also given are the contributions from  all methyl groups.}
\label{fig:IR3000}
\end{figure}
\begin{figure}
\includegraphics[width=\columnwidth]{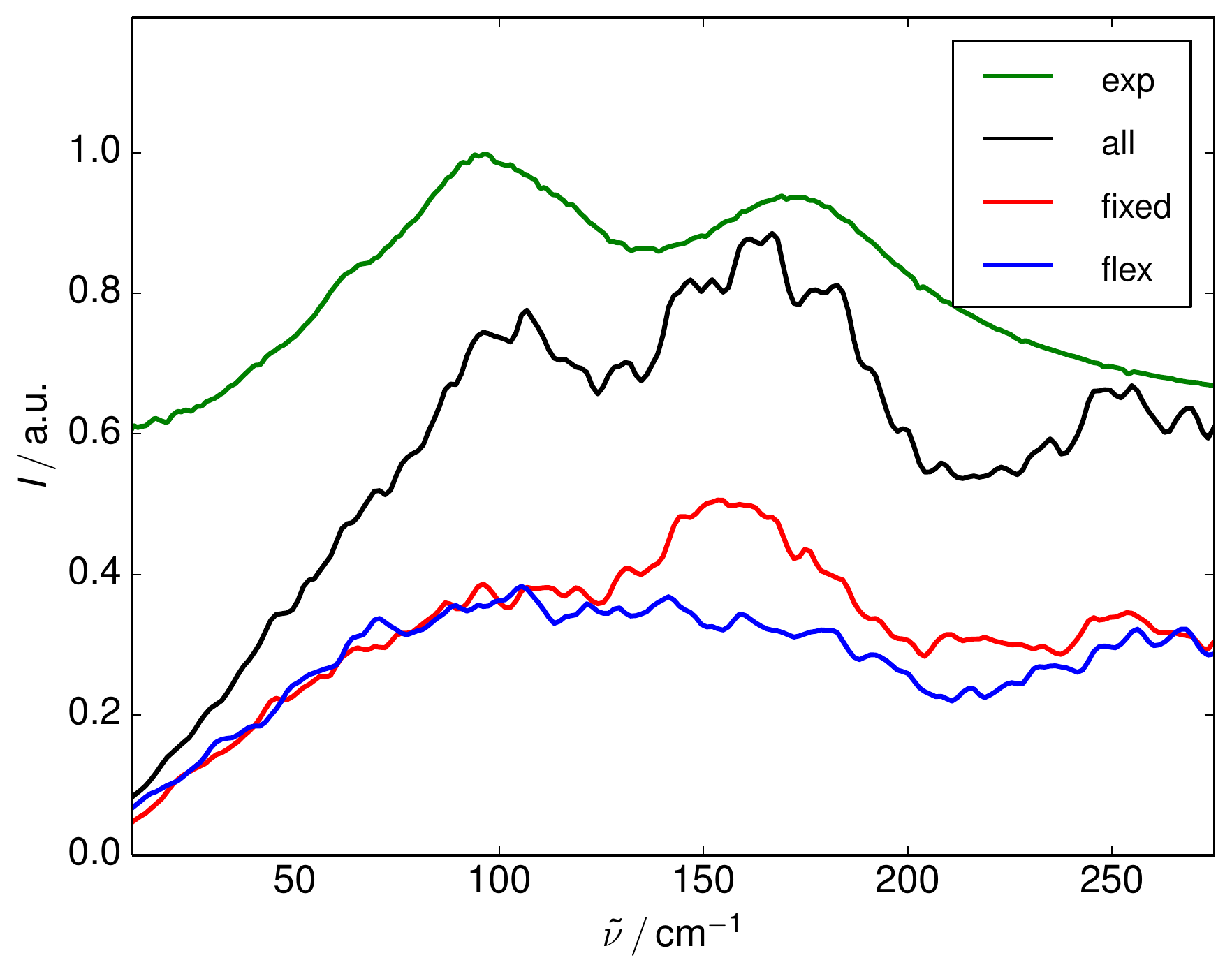}
\caption{DFTB far IR spectra according to Eq.~\eqref{eq:IR} as well as experimental results from Fumino \textit{et al.} \cite{fumino12_6236}. The separate contributions of fixed and flexible NH$\cdots$O HBs are shown as well.}
\label{fig:FIR}
\end{figure}

The spectral signatures of the vibrations in the CH- and NH-stretching range  around 3000~$\mathrm{cm^{-1}}$ are highlighted in  Fig.~\ref{fig:IR3000}. The spectrum consists mainly of two peaks, around 2950~$\mathrm{cm^{-1}}$ and 3070~$\mathrm{cm^{-1}}$. To disentangle the contributions of   ion pairs differing in their HB strength, spectra for `fixed' and `flexible' ion pairs are shown separately. Apparently, the spectra from the fixed pairs have more intensity for the lower-frequency peak of the  spectrum.  This points to the expected stronger HB present for the fixed pairs. To strengthen this argument further the spectrum according to the NH-bond vibration only has been calculated for the fixed and flexible ion pairs. As can be seen from  Fig.~\ref{fig:IR3000} this yields rather broad spectra, with a maximum at 3000~$\mathrm{cm^{-1}}$ and 2950~$\mathrm{cm^{-1}}$ for the flexible and fixed case, respectively, in accord with the assumed HB strength.
 
In addition to NH- also CH-stretching vibrations are contributing to this frequency range. The contributions from the methyl groups only are shown in Fig.~\ref{fig:IR3000}, which yields the conclusion that this spectral range is dominated by the absorption of the ethyl (around 2950~$\mathrm{cm^{-1}}$) and methyl (around  3050~$\mathrm{cm^{-1}}$) groups. In passing we note that in case of a crystal made of the the same cation but a bromine anion, Mondal \textit{et al.}~\cite{mondal15_1994} using Car-Parinello MD simulations found a double peak structure in the vibrational density of states, which was attributed to NH- and CH-stretching vibrations.

Finally, we consider the far IR region shaped by intermolecular HB interactions. The results from the DFTB simulation are plotted in Fig.~\ref{fig:FIR}, for the full system as well as for the fixed and flexible configurations.  Two bands can be identified. One has a maximum at 166~$\mathrm{cm^{-1}}$ and is dominated by contributions from fixed ion pairs. The other one which has a maximum at about 104~$\mathrm{cm^{-1}}$ has considerable contributions from the flexible pairs. Hence, for the stronger H-bonded pairs the stronger red shift in the NH-stretching range comes along with a blue-shifted absorption in the far IR range of the HB mode.

The H-bonding contributions to this spectral region was analyzed in detail through measurement from Fumino \textit{et al.} \cite{fumino12_6236}, who showed that H-bonding adds an extra blue shifted band as compared to contributions from  dispersion interaction. Experimental data for the related trimethylammonium nitrate are also shown in Fig.~\ref{fig:FIR}.  Fumino \textit{et al.} assigned the blue-shifted peak  at 171~$\mathrm{cm^{-1}}$ to a H-bonding mode, whereas the lower frequency peak at 96~$\mathrm{cm^{-1}}$ is assigned to unspecific libration modes, which could also involve the NH-bond for the flexible case. Overall the agreement is very good, even though the simulations were performed with ethyl and not methyl side chains and for a different phase (solid phase at 300~K). The longer alkyl chain leads to an inductive effect and a higher electron density at the nitrogen. This causes a  weakening of the HB, which was analyzed for n-alkylammonium methylsulfonate in Ref.~\citenum{fumino12_6236}. 
The HB strength impacts the frequency offset between the two bands, where a stronger HB leads to a larger blue shift of the H-bonding modes. Indeed, as expected we find the methylammonium nitrate peaks separated by 75~$\mathrm{cm^{-1}}$, while (the presumably weaker H-bonding) ethylamonium peaks are separated by 62~$\mathrm{cm^{-1}}$. 

The FF MD approach gives no IR signal below 200~$\mathrm{cm^{-1}}$.  As a side remark we note that  Sunda \textit{et al.}\cite{Sunda15_4625} reported FF MD simulations of power spectra (vibrational density of states) for triethylammonium triflate, which showed only one band in region of 100 to 200~$\mathrm{cm^{-1}}$.
\section{Summary}
\label{sec:sum}
This contribution has served two goals: First, to scrutinize the performance of DFTB by comparison with other methods. Second, to characterize the HB dynamics and IR spectroscopy for the specific case of tEAN. Overall, DFTB performed rather well in describing the potential energy surface related to HB motion in tEAN. In gas as well as liquid phase, DFTB potential energy curves along the NH-bond coordinate have been rather close to high-level CCSD(T) data in the thermally accessible range. Further, comparison with DFT results for a hexamer cluster yielded a favorable agreement in bond distance correlations, which supports the conclusion that DFTB reliably describes the energetics of  a wider range of HB geometries.

Based on these encouraging findings, DFTB liquid phase MD simulations have been performed and compared to results obtained using an established force field. Geometric correlations between bond distances of the HB were analyzed within an approach widely used by Limbach and coworkers, which is based on Pauling's bond order theory. The distribution of HB geometries indicated weak correlations, thus pointing to a medium to weak HB. Interestingly the bond order parameters are not much different from those obtained for NHO HBs of non-ionic systems. In contrast FF simulations did not show any signature of correlations.

Concerning IR spectra, FF and DFTB results show rather pronounced differences, which can be traced to the poor description of the potential energy surface provided by the force field. DFTB spectra have been disentangled according to contributions from strongly and flexibly H-bonded ion pairs. In the spectral range of the NH-stretching vibrations this yielded a bimodal distribution of intensities according to the HB strength. In the total absorption spectrum the NH-stretching features are hidden underneath the ethyl and methyl group absorption. An even more pronounced bimodal behavior has been found in the far IR region. Here, the almost quantitative agreement with experimental data lends strong support for the DFTB method.

\begin{acknowledgments}
We gratefully acknowledge financial support by the  Deutsche Forschungsgemeinschaft through the SFB 652.
\end{acknowledgments}

%

 %
 
\end{document}